\documentclass[%
  aps,          % American Physical Society style
  prl,          % Physical Review Letters
  twocolumn,    % PRL uses two-column layout
  showpacs,     % show PACS numbers (optional but encouraged)
  preprintnumbers,
  superscriptaddress, % allows multiple affiliations per author
  longbibliography,   % include reference titles (PRL strongly encourages this)
]{revtex4-2}
\usepackage{amsmath}        % math environments
\usepackage{amssymb}        % extra math symbols
\usepackage{graphicx}       % \includegraphics
\usepackage{dcolumn}        % align table columns on decimal point
\usepackage{bm}             % bold math (\bm{})
\usepackage{hyperref}       % clickable links in PDF
\usepackage{xcolor}         % color support (useful for draft notes)
\usepackage{braket}         % Dirac bra-ket notation
\usepackage{comment}        % \begin{comment} environment
\usepackage{mathtools}      % extends amsmath
\usepackage{tikz}           % diagrams
\usepackage{subcaption}     % subfigure environment
\usepackage{float}          % [H] float placement specifier
\usepackage{enumitem}       % itemize/enumerate with custom options
\usepackage{amsthm}         % theorem-like environments
\newtheorem{defn}{Definition}
\newtheorem{thm}{Theorem}
\DeclarePairedDelimiter\abs{\lvert}{\rvert}

\newcommand{\eq}[1]{\begin{equation}#1\end{equation}}

\newcommand{\eqsp}[1]{\begin{equation}\begin{split}#1\end{split}\end{equation}}
\begin{document}
\preprint{MIT-CTP/6069}
% ── Title ─────────────────────────────────────────────────
\title{The \texorpdfstring{$\boldsymbol{\alpha}$}{a}-states of a string worldsheet}

% ── Authors and affiliations ──────────────────────────────
% Use \author + \affiliation pairs; share affiliations via \altaffiliation
\author{Suzanne Bintanja}
\email{sbintanja@physics.ucla.edu}
\affiliation{Mani L. Bhaumik Institute for Theoretical Physics, Department of Physics and Astronomy, University of California Los Angeles, Los Angeles, CA 90095, USA}
\affiliation{CERN, Theory Division,
Geneva 23, CH-1211, Switzerland}

\author{Elliott Gesteau}
\email{egesteau@mit.edu}
\affiliation{Center for Theoretical Physics -- a Leinweber Institute, Massachusetts Institute of Technology, \\Cambridge, MA 02139, USA}
\affiliation{Center of Mathematical Sciences and Applications, Harvard University, Cambridge MA, 02138, USA}

% ── Date (leave blank to suppress) ───────────────────────
%\date{\today}

% ── Abstract (150 words max for PRL) ─────────────────────
\begin{abstract}
The worldsheet description of string amplitudes can be reinterpreted as a baby universe field theory. Under this reinterpretation, we determine the $\alpha$-states of a topological string theory: the Hurwitz worldsheet. We find that their relation to the Hartle--Hawking state encodes foundational results in the asymptotic representation theory of the symmetric group. In the holographic dual, our results characterize the pseudorandom nature of an ensemble of cluster-decomposing correlators.
\end{abstract}

% ── Keywords ─────────────────────────────────────────────
%\keywords{keyword one, keyword two, keyword three}

\maketitle

\section*{Introduction}

The idea of a third-quantized baby universe field theory (BUFT), originally proposed by Coleman, Giddings, and Strominger \cite{Coleman:1988cy,Giddings:1988cx,Giddings:1988wv}, owes its modern formulation to Marolf and Maxfield \cite{MarMax20}. In this formulation, many-universe amplitudes are written as expectation values of a commutative algebra of universe field operators in the Hartle--Hawking state of a baby universe Hilbert space $\mathcal{H}_{BU}$. 

A central aspect of BUFT is that there exists a basis of ``$\alpha$-states" in $\mathcal{H}_{BU}$ that diagonalizes the algebra of universe field operators. When rewritten in this basis, many-universe amplitudes admit another interpretation: they compute statistical moments of expectation values of universe field operators in an ensemble of $\alpha$-states. This ensemble is obtained by independent projective measurements of the Hartle--Hawking state \cite{MarMax20}.

In a given $\alpha$-state, expectation values of universe field operators factorize, which is a property required for holographic theories \cite{MalMao04}. For a single UV-complete holographic theory, the associated BUFT contains a single $\alpha$-state and hence $\mathcal{H}_{BU}$ is one-dimensional \cite{Pen19,McNVaf20,UsaZha24,UsaWan24,HarUsa25,AbdAnt25,EngGes25a,Ges25}. On the other hand, BUFT theories with more than one $\alpha$-state do not possess a standard holographic dual, and in some examples, the basis of $\alpha$-states has been identified with an ensemble of holographic boundary theories \cite{Saad:2019lba,MarMax20}. 

The worldsheet of string theory, when interpreted as a closed universe made out of a string, admits a BUFT formulation \cite{McNVaf20,CasMar21,Post:2022dfi,Harlow:2026hky,Zha26}, where the analogs of universe field operators insert vertex operators. Because string amplitudes do not factorize, the worldsheet is not holographic as a theory of quantum gravity, and its baby universe Hilbert space can be more than one-dimensional \footnote{This does not mean that a dual gauge theory does not factorize. In fact, the dual TQFT relevant to this letter does factorize. In this case, the $\alpha$-states should be identified with different states of the dual TQFT, see e.g. \cite{Sharpe:2023lfk}}. The BUFT formulation then allows us to interpret string amplitudes as statistical moments of vertex operators in an ensemble of $\alpha$-states \cite{McNVaf20,CasMar21,Post:2022dfi,Harlow:2026hky}. What is this ensemble, and what statistical information do string amplitudes give us about it?  

In this Letter, we initiate a study of the $\alpha$-states of the string worldsheet. The worldsheet theory we consider is the Hurwitz worldsheet \cite{okounkov2002gromovwittentheoryhurwitztheory,Benizri:2024mpx,Komatsu:2025sqo}, which can be realized as a topological string theory on AdS${_3}$ \cite{Knightontoappear}. In this theory, we show that the $\alpha$-states, seen as linear functionals on the vertex operators, are given by suitably normalized characters of the irreducible representations $R$ of the symmetric groups $S_N$ for all values of $N$. The probability to measure a given $\alpha$-state $\ket{R}$ in the Hartle--Hawking state is equal to the ``Poissonized" Plancherel measure evaluated on $R$. In the boundary dual of our worldsheet theory, known as Hurwitz TQFT, the $\alpha$-states can be interpreted as cluster-decomposing states.

Using the expansion of the string amplitudes in the string coupling $g_s$, we show that expectation values of suitably normalized string insertions in the ensemble of $\alpha$-states satisfy a central limit theorem in the weak-coupling limit: they become \textit{pseudorandom}. That is, in the limit $g_s\rightarrow 0$, these expectation values cannot be distinguished from independent Gaussian random variables by any fixed number of measurements. In the dual Hurwitz TQFT, our result implies that expectation values of operators in cluster-decomposing states become pseudorandom in the large-$N$ limit. This gives a natural physical interpretation to Kerov's central limit theorem \cite{ivanov1999algebra,ivanov2001algebra,Ker00,ivanov2002kerov}, a foundational result in the mathematical field of asymptotic representation theory.

More broadly, determining the $\alpha$-states of the worldsheet allows one to reverse-engineer an ensemble of quantities that become pseudorandom as $g_s\rightarrow 0$. Already in the case at hand, these quantities have a precise interpretation in terms of the dual TQFT, and connect to deep mathematical structures. We see this as a hint that understanding the $\alpha$-states of the string worldsheet may provide new insights into the structure of string theory.

\section*{Review of baby universe field theory}

We briefly review the Marolf--Maxfield formulation of BUFT \cite{MarMax20}. First, consider an abelian algebra of observables $\hat{Z}(J)$, whose role is to insert a boundary condition $J$ in the Euclidean past (this boundary condition labels both the boundary manifold and sources). The algebra generated by the $\hat{Z}(J)$ is equipped with an involution $\ast$, which CPT conjugates the boundary condition:
\begin{align}
\hat{Z}(J)^\ast=\hat{Z}(J^\ast)\,.
\end{align}

We then define a formal vector space generated by vectors of the form 
\begin{align}\ket{Z(J_1)\dots Z(J_k)}=\hat{Z}(J_1)\dots\hat{Z}(J_k)\ket{\mathrm{HH}}\,,\end{align}
where $\ket{\mathrm{HH}}$ is an abstract reference vector which should be understood as the Hartle--Hawking state, and equip it with the inner product
\begin{align}\begin{split}
&\bra{\mathrm{HH}}\hat{Z}(J_1)^\ast\dots\hat{Z}(J_{k})^\ast\hat{Z}(J^\prime_{k^\prime})\dots\hat{Z}(J^\prime_1)\ket{\mathrm{HH}}\\
&\hspace{80pt}:=\mathrm{GPI}(J^\ast_1,\dots,J^\ast_k,J^{\prime}_{k^\prime},\dots, J^{\prime}_1)\,.
\label{eq:gpi}\end{split}
\end{align}
Here, the right-hand side denotes the gravitational path integral computed with boundary conditions $J_1^\ast,\dots,J_k^\ast,J^\prime_1,\dots, J^\prime_k$. When this inner product is positive semidefinite, we can then quotient the above vector space by the null states of the inner product to obtain a baby universe Hilbert space $\mathcal{H}_{BU}$.

Because the algebra generated by the $\hat{Z}(J)$ and the $\hat{Z}(J)^\ast$ is commutative, the operators can all be simultaneously diagonalized in a special basis of the baby universe Hilbert space, commonly known as the basis of $\alpha$-states $\ket{\alpha}$ \cite{MarMax20}. These $\alpha$-states have the property that they factorize on baby universe observables
\begin{align}
\bra{\alpha}\hat{Z}(J_1)\hat{Z}(J_2)\ket{\alpha}=\bra{\alpha}\hat{Z}(J_1)\ket{\alpha}\bra{\alpha}\hat{Z}(J_2)\ket{\alpha}\,.
\end{align}
This factorization property is required if we interpret the path integral of the right-hand side of \eqref{eq:gpi} as computing a path integral in a UV-complete holographic theory with disconnected boundary conditions. This means that in holographic theories, the Hartle--Hawking state coincides with an $\alpha$-state and the baby universe Hilbert space is one-dimensional \cite{Pen19,McNVaf20,UsaZha24,UsaWan24,HarUsa25,AbdAnt25,EngGes25a,Ges25}.

Correlation functions in the Hartle--Hawking state can be expressed as
\begin{align}\begin{split}
\frac{1}{\aleph}\bra{\mathrm{HH}}\hat{Z}(J_1)^\ast\dots\hat{Z}(J_{k})^\ast\hat{Z}(J^\prime_{k^\prime})\dots\hat{Z}(J^\prime_{1})\ket{\mathrm{HH}}\\
=\int d\alpha\, \mu(\alpha)\bra{\alpha}\hat{Z}(J_1^\ast)\ket{\alpha}\dots\bra{\alpha}\hat{Z}(J^\prime_{k^\prime})\ket{\alpha}\,,
\label{eq:HHalpha}
\end{split}
\end{align}
where $\aleph=\braket{\mathrm{HH}|\mathrm{HH}}$ and $\mu(\alpha)$ is a probability measure on the space of $\alpha$-states.
Eq. \eqref{eq:HHalpha} has a statistical interpretation. Suppose we prepare a very large number of copies of the Hartle--Hawking state and measure each of them in the $\alpha$-basis. The statistical moments of expectation values of the universe field operators $\hat{Z}(J)$ in the resulting collection of states are given by Eq. \eqref{eq:HHalpha}. As mentioned above, the string worldsheet can itself be interpreted as a theory of BUFT, where the universe field operators correspond to vertex operator insertions. In the case of the string worldsheet, what statistical information do the moments of \eqref{eq:HHalpha} give us access to?

\section*{The Hurwitz worldsheet and its \texorpdfstring{$\boldsymbol{\alpha}$}{a}-states}

\subsection*{The worldsheet theory}

The worldsheet theory we consider in this Letter is topological and can be obtained as a limit of string theory on AdS$_3$ \cite{Knightontoappear,Knightoncommunication} (see also \cite{okounkov2002gromovwittentheoryhurwitztheory,Benizri:2024mpx,Komatsu:2025sqo} for related discussions). It has vertex operators labelled by the winding numbers of the associated closed strings. The worldsheet correlators, equal to the closed string amplitudes, are computed with integrals over moduli space. Such integrals are delta-function localized on string moduli space to configurations that allow for a (possibly disconnected) covering map of the $S^2$ boundary of AdS$_{3}$ by the worldsheet \cite{Eberhardt:2019ywk,Dei:2020zui,Knighton:2020kuh,Komatsu:2025sqo}. Each covering surface contributes the simple weight $g_s^{-2N}$, with $N$ the number of sheets in the covering, appropriately weighted by a symmetry factor \footnote{The symmetry factor is equal to the inverse of the order of the automorphism group of the covering map.}\footnote{See \cite{Aharony:2024fid} for a discussion on the choices of normalization for the vertex operators.}.

To obtain a BUFT formulation of the theory, we introduce universe field operators as the string insertion operators $\hat{Z}(\mu)$, where $\mu$ denotes the integer labeling the worldsheet vertex operators. For simplicity, we will mostly consider single-string insertions; multi-string insertions are instead labelled by integer partitions with multiple elements. In our setup, the involution $\ast$ on the algebra generated by the $\hat{Z}(\mu)$ is trivial (see \cite{CasMar21} for details). The expectation value of $\hat{Z}(\mu_1)\dots\hat{Z}(\mu_k)$ in the Hartle--Hawking state is then by definition equal to the amplitude obtained by evaluating the worldsheet path integral, which corresponds to the integral over string moduli space
\begin{align}
\bra{\mathrm{HH}}\hat{Z}(\mu_1)\dots\hat{Z}(\mu_k)\ket{\mathrm{HH}}:=\mathcal{A}_{\mu_1,\dots,\mu_k}\,.
\end{align}
Because the worldsheet path integral localizes to covering surfaces of the sphere, the (normalized) amplitude is given by
\eqsp{\label{eq:gcampl}
&\frac{1}{\aleph}\bra{\mathrm{HH}}\hat{Z}(\mu_1)\dots\hat{Z}(\mu_k)\ket{\mathrm{HH}}
\\
&\hspace{80pt}=e^{-p}\sum_{N=0}^\infty \frac{p^N}{N!} H^{\bullet N}_{\mu_1,\dots,\mu_k}\,.
}
Here $\aleph=\langle\mathrm{HH}\vert \mathrm{HH}\rangle=e^p$ is the squared norm of the Hartle--Hawking state, the string coupling is related to the grand canonical potential via $g_s^{-2}\propto p$ \cite{Aharony:2024fid}, and $H^{\bullet N}_{\mu_1,\dots,\mu_k}$ is a \textit{Hurwitz number}. Hurwitz numbers, when divided by $N!$, count the number of $N$-sheeted ramified covering surfaces, %of the punctured $S^2$ that the worldsheet path integral localizes to, 
weighted by their symmetry factors, with ramification profiles $\mu_1,\dots,\mu_k$. These are now interpreted as single-cycle conjugacy classes of the symmetric group. Explicitly, the Hurwitz numbers are given by
\eq{\label{eq:ampl}
H^{\bullet N}_{\mu_1,\dots,\mu_k}\coloneqq\Bigg\vert\left\{g_i\in\mu_i : \prod_{i=1}^k g_i=\text{id}\right\}\Bigg\vert\,.
}

\subsection*{The \texorpdfstring{$\boldsymbol{\alpha}$}{a}-states}

The crucial formula for the purposes of understanding the $\alpha$-states of our string worldsheet expresses the aforementioned Hurwitz numbers of the sphere \eqref{eq:ampl} as sums over irreducible representations of the symmetric group $S_N$ (some relevant properties of the representation theory of the symmetric group are collected in the appendix) \cite{burnside1911}
\eq{
H^{\bullet N}_{\mu_1,\dots,\mu_k}=\hspace{-10pt}\sum_{R\in\mathrm{Irr}S_N}\frac{(\mathrm{dim}R)^{2}}{N!}\prod_{i=1}^k\abs{\mu_i}\frac{\chi_R(\mu_i)}{\dim R}\,.
\label{eq:Hurwitz}
}
Here $\abs{\mu}$ denotes the number of elements in the conjugacy class $\mu$, and $\chi_R(\mu)$ is the irreducible character of the irrep $R$ evaluated on the  $S_N$ conjugacy class $\mu$.

Irreducible representations of the symmetric group are labelled by Young diagrams with $N$ boxes (see the appendix for details). Hence, we can rewrite \eqref{eq:gcampl} as a sum over Young diagrams (YD) with any number of boxes, so as to obtain
\begin{align}
\begin{split}
\label{eq:stringamp}
&\frac{1}{\aleph}\bra{\mathrm{HH}}\hat{Z}(\mu_1)\dots\hat{Z}(\mu_k)\ket{\mathrm{HH}}\\
&\hspace{20pt}=\sum_{R\in\rm{YD}}e^{-p}p^{\abs{R}}\left(\frac{\mathrm{dim}R}{\abs{R}!}\right)^{2}\prod_{i=1}^k\abs{\mu_i}\frac{\chi_R(\mu_i)}{\dim R}\,,
\end{split}
\end{align}
where $|R|$ denotes the number of boxes of the Young diagram $R$.

It is striking that Eq. \eqref{eq:stringamp} takes the exact form of Eq. \eqref{eq:HHalpha} describing a BUFT correlation function in the Hartle--Hawking state! In particular, it tells us all there is to know about the baby universe field theory of the Hurwitz worldsheet:
\begin{itemize}[leftmargin=10pt]
\item[]\textbf{Baby universe Hilbert space.} An orthonormal basis of the baby universe Hilbert space of the worldsheet $\mathcal{H}_{BU}$ is labeled by Young diagrams of all sizes. 

\item[]\textbf{$\boldsymbol{\alpha}$-states.} The $\alpha$-state basis is given by the states $\ket{R}$, which label irreducible representations of $S_{\abs{R}}$ of any size $\abs{R}$. The expectation values of the closed string insertion operators $\hat{Z}(\mu)$ in $\ket{R}$ are given by normalized characters of $S_{\abs{R}}$
\begin{align}
\bra{R}\hat{Z}(\mu)\ket{R}
=\frac{\abs{\mu}\chi_R(\mu)}{\dim R}\,.\label{eq:alphastates}\end{align}

\item[]\textbf{Hartle--Hawking state.} The overlap of the Hartle--Hawking state $\ket{\mathrm{HH}}$ with an $\alpha$-state $\ket{R}$ of size $\abs{R}$ is given by the ``Poissonized" Plancherel measure \cite{Bogachev2006Central}
\begin{align}
\frac{1}{\aleph}|\langle\mathrm{HH}\vert R\rangle|^2=e^{-p}p^{\abs{R}}\left(\frac{\dim R}{\abs{R}!}\right)^2\,,
\label{eq:measure}
\end{align}
with $\aleph=\langle \text{HH}|\text{HH}\rangle=e^p$. Using the fact that the Plancherel measure is a probability measure on Young diagrams of size $\abs{R}$ (see the appendix), it is easy to see that \eqref{eq:measure} is a probability measure on the space of all Young diagrams. 
\end{itemize}

\subsection{Boundary interpretation and the microcanonical ensemble}

One of the distinguishing features of the Hurwitz worldsheet, and more generally of the worldsheet of the tensionless string in AdS$_3$/CFT$_2$, is that its amplitudes are exactly equal to the correlators of a symmetric product orbifold CFT in the grand-canonical ensemble \cite{Knighton:2024pqh}. In our case, as established in \cite{okounkov2002gromovwittentheoryhurwitztheory,Benizri:2024mpx,Komatsu:2025sqo,Knightontoappear}, the dual theory is Hurwitz TQFT. Its Hilbert space $\mathcal{H}_N$ is spanned by conjugacy classes, or equivalently, irreducible representations, of $S_N$. Moreover, its structure constants are given by the Hurwitz numbers of the three-punctured sphere $H_{\mu_1,\mu_2,\mu_3}^{\bullet N}$ \cite{Dijkgraaf:1989pz,Dijkgraaf1995} (for a more recent discussion see \cite{Li:2020zwo}). We have the exact isomorphism of Hilbert spaces
\begin{align}
\mathcal{H}_{BU}\cong\bigoplus_{N=0}^\infty\mathcal{H}_N,
\end{align}
where $\mathcal{H}_N$ is the Hilbert space of the Hurwitz TQFT with gauge group $S_N$. The Hurwitz TQFT counterpart of an $\alpha$-state is a state in which expectation values of operators factorize, i.e., a cluster-decomposing state \cite{nextpaper}.

If we want a prescription to modify the worldsheet theory presented above to go to the microcanonical ensemble, i.e. to obtain a state $\ket{\mathrm{HH}_N}$ in the baby universe Hilbert space $\mathcal{H}_{BU}$ with support solely on $\mathcal{H}_N$, we can choose to only sum over worldsheets that are covers of the sphere with exactly $N$ sheets. In such a state, 
\eqsp{
&\frac{1}{\aleph_N}\langle\mathrm{HH}_N|\hat{Z}(\mu_1)\dots\hat{Z}(\mu_k)|\mathrm{HH}_N\rangle\\
&\hspace{40pt}=\sum_{R\in\mathrm{Irr}\,S_N}\frac{(\dim R)^2}{N!}\prod_{i=1}^k\abs{\mu_i}\frac{\chi_R(\mu_i)}{\dim R}\, ,
\label{eq:microc}
}
with $\aleph_N=\frac{p^N}{N!}$. For string insertions with $\mathcal{O}(1)$ winding, the leading order of the expansion in $p$ of \eqref{eq:stringamp} matches the leading order of the expansion in $N$ of \eqref{eq:microc} \cite{Aharony:2024fid}; this is the statement of equivalence of the microcanonical and grand-canonical ensembles in the thermodynamic limit.

It is interesting to note that a very similar rule to the one we described here is used in \cite{MarMax20} to modify the gravitational path integral to compute correlators of universe field operators in an $\alpha$-state. There, the rule is to restrict to a sum over spacetimes with a fixed number of connected components; in our setup, the rule is to restrict to worldsheets with a fixed number of sheets. Note that, as emphasized in \cite{GesMar24}, these kinds of restrictions are nonlocal in nature. Because the theory under consideration is more complicated than the BUFT of \cite{MarMax20}, the rule is not quite enough to construct a single $\alpha$-state. Nonetheless, it produces a state with support on only finitely many $\alpha$-states. It would be interesting to understand how to systematically modify the rules of the sum over worldsheets to prepare an $\alpha$-state in $\mathcal{H}_{BU}$, perhaps using the methods developed in \cite{GesMar24}. 

\section*{Kerov's central limit theorem from the worldsheet}

The correlations in the Hartle--Hawking state \eqref{eq:stringamp} can be reinterpreted in probabilistic terms. To do so, we interpret the functions $\bra{R}\hat{Z}(\mu)\ket{R}$ as random functions of the Young diagram $R$. Eq. \eqref{eq:stringamp} then expresses the joint moments of these random variables over an ensemble of Young diagrams. The measure of a Young diagram $R$ in this ensemble is the probability of measuring $\ket{R}$ in the Hartle--Hawking state.
 
The dominant contribution to the BUFT amplitude $\bra{\mathrm{HH}}\hat{Z}(\mu_1)\dots\hat{Z}(\mu_k)\ket{\mathrm{HH}}$ comes from disconnected worldsheets that connect the insertions $\mu_i$ pairwise when possible \footnote{Related to this fact is the string genus expansion \cite{Aharony:2024fid}.}. All other contributions are suppressed by powers of $p$ (or equivalently $g_s$). Such leading contributions scale like a positive power of $p$. In particular, in the large-$p$ limit, for $\mu_1,\dots,\mu_k$ mutually distinct single-cycle conjugacy classes (whose length we also denote by $\mu_i$), we have
\eqsp{
&\underset{p\rightarrow\infty}{\mathrm{lim}}\frac{1}{\aleph}p^{-\sum_{i=1}^k\frac{\mu_in_i}{2}}\bra{\mathrm{HH}}\prod_{i=1}^k\hat{Z}(\mu_i)^{n_i}\ket{\mathrm{HH}}\\
&\hspace{40pt}=\prod_{i=1}^k\underset{p\rightarrow\infty}{\mathrm{lim}}\frac{1}{\aleph}p^{-\frac{\mu_in_i}{2}}\bra{\mathrm{HH}}\hat{Z}(\mu_i)^{n_i}\ket{\mathrm{HH}}\,,
}
with
\begin{align}
\begin{split}
&\lim_{p\rightarrow\infty}\frac{1}{\aleph}\left(\frac{\mu_i}{p^{\mu_i}}\right)^{\frac{n_i}{2}}\langle\mathrm{HH}|\hat{Z}(\mu_i)^{n_i}|\mathrm{HH}\rangle \\
&\hspace{80pt}=
    \begin{cases}
        0 & n_i \text{ odd}, \\[6pt]%\text{if } n_i \text{ is odd,} \\[6pt]
        (n_i-1)!! & n_i \text{ even.}%\text{if } n_i \text{ is even.}
    \end{cases}
    \end{split}
\end{align}
Here the factors of $\mu_i$ are the symmetry factors of the pairwise connecting covering surfaces. In these equations, we recognize the joint moments of $k$ mutually independent Gaussian random variables.

In probability theory, under minor technical assumptions satisfied here (see \cite{ivanov2002kerov} for details), the convergence of all the joint moments of random functions towards the joint moments of a given set of random variables implies convergence in distribution: for all $x_1,\dots,x_k\in\mathbb{R}$,
\begin{align}\label{eq:gauss}
&\nonumber\hspace{0pt} \lim_{p\rightarrow\infty}p^{-\sum_{i=1}^k\frac{\mu_i}{2}}\mathbb{P}(\bra{R}\hat{Z}(\mu_1)\ket{R}\leq x_1,\dots,\bra{R}\hat{Z}(\mu_k)\ket{R}\leq x_k)\\
&\hspace{40pt}=\mathbb{P}(\mathcal{N}_1\leq x_1,\dots,\mathcal{N}_k\leq x_k)\,,
\end{align}
where the $\mathcal{N}_i$ are independent Gaussian random variables with variance $\mu_i^{-1}$. Here, the probability is computed according to the measure \eqref{eq:measure}, i.e., the probability to measure an $\alpha$-state corresponding to a Young diagram $R$ in the Hartle--Hawking state.
Recalling that by \eqref{eq:alphastates} the random functions $\bra{R}\hat{Z}(\sigma_i)\ket{R}$ are (normalized) characters of irreducible representations of symmetric groups labeled by Young diagrams, in mathematical terms, this means that these characters follow a central limit theorem when sampled over the Poissonized Plancherel measure! It turns out that this result, known as Kerov's central limit theorem \cite{ivanov1999algebra,ivanov2001algebra}, is one of the cornerstones of the mathematical field of asymptotic representation theory. Discussions of Kerov's central limit theorem are usually done in the microcanonical ensemble, whereas the genus expansion of the string worldsheet is valid in the grand-canonical ensemble. However, to leading order, the ensembles are equivalent so that a central limit theorem for $\alpha$-states in one implies a central limit theorem in the other. Proving limit theorems in probability theory by going to the grand-canonical ensemble is sometimes referred to as Poissonization \cite{Bogachev2006Central,CheJun25}.

Eq. \eqref{eq:gauss} provides a physical interpretation of Kerov's central limit theorem. Suppose that one performs a projective measurement of the $\alpha$-parameter on many copies of the Hartle--Hawking state of the Hurwitz worldsheet. Then the distribution of expectation values of string operators with $\mathcal{O}(1)$ winding in the resulting $\alpha$-states becomes \textit{pseudorandom}, i.e., statistically indistinguishable from independent Gaussian statistics, as $g_s\rightarrow 0$.

There is an equivalent interpretation of Kerov's central limit theorem in the dual Hurwitz TQFT. The statement is that expectation values of operators in cluster-decomposing states become pseudorandom in the large-$N$ limit. A detailed treatment of the mechanism in Hurwitz TQFT will be presented in forthcoming work \cite{nextpaper}.

It is interesting to contrast our result with what happens in the baby universe field theory of the semiclassical gravitational path integral. Indeed, there is also a connection between the gravitational path integral and the statistical properties of its $\alpha$-states. In particular, at least in low spacetime dimension, leading contributions to the path integral are expected to encode the joint moments of an ensemble of boundary CFT data in the semiclassical limit \cite{BeldeB20,Belin:2021ryy,Chandra:2022bqq,deBoer:2023vsm,Belin:2023efa,deBoer:2024mqg,Jafferis:2025vyp,Belin:2026pko}. There, subleading corrections are expected to be exponentially suppressed in $1/G_N$, as they come from subleading gravitational saddles. On the other hand, in our setup, subleading corrections are only power-law suppressed in $g_s$.

\section*{Discussion}

In this Letter, we derived the BUFT formulation of the Hurwitz worldsheet theory and computed its $\alpha$-states. We found that, when sampled over independent measurements of the Hartle--Hawking state, expectation values of vertex operators in $\alpha$-states follow a central limit theorem as $g_s\rightarrow 0$. This provides a physical interpretation of Kerov's central limit theorem, a foundational result in asymptotic representation theory. In the dual topological field theory, this translates to the large-$N$ pseudorandomness of operator expectation values in cluster-decomposing states.

We believe that the BUFT formulation of the worldsheet, and its connection to asymptotic representation theory presented in this work, can lead to new insights into the probabilistic interpretation of the string worldsheet, pseudorandomness in topological field theory, and maybe even pure mathematics. Future directions, some of which will be investigated in forthcoming work \cite{nextpaper}, include, in no particular order:
\begin{itemize}[label={$\diamondsuit$}, leftmargin=1.2em]
\item Generalizing our results to other worldsheet theories. 
\item Analyzing the non-Gaussian corrections to the $\alpha$-state ensemble coming from subleading terms in the genus expansion.
\item Understanding the relationship between the pseudorandomness of characters of the symmetric group and interface theory in TQFT, see e.g. \cite{Gutperle:2024vyp,Knighton:2024noc,Benjamin:2025knd,Harris:2025wak}.
\item Finding a bulk procedure to construct an $\alpha$-state of the worldsheet, perhaps following the analogy of \cite{GesMar24} with the BPHZ algorithm from perturbative renormalization.
\item Understanding the role of worldsheet $\alpha$-states in target space, and their relation to branes and open strings.
\end{itemize}

\section*{Acknowledgments}
We would like to thank Alex Belin, Kasia Budzik, Gabriele Di Ubaldo, Netta Engelhardt, Matthias Gaberdiel, Diego García-Sepúlveda, Daniel Harlow, Bob Knighton, Ji Hoon Lee, Hong Liu, Jacob McNamara, Shu-Heng Shao, Adar Sharon, Dikshant Rathore, Wayne Weng, Xingyang Yu, and Ying Zhao for helpful discussions. We also thank the organizers and participants of the LITP workshop Quantum Black Holes and the GGI workshop Pathways to Quantum Black Holes for fruitful discussions and support. The work of SB is supported by the Mani L. Bhaumik Institute for Theoretical Physics. The work of EG is supported in part by the Department of Energy under Early Career Award DE-SC0021886, by the Heising-Simons Foundation under grant no. 2023-4430, and by the Moore Foundation via the Black Hole Initiative.

\appendix

\section{Appendix}\label{app:young}

In this appendix, we discuss relevant background on the representation theory of the symmetric group. 
\begin{itemize}[leftmargin=10pt]
\item[]\textbf{Young diagrams.} Irreducible representations of the symmetric group $S_N$ are, just as conjugacy classes, labelled by partitions of $N$. A very useful way to represent these partitions of $N$ is through Young diagrams, which we will take to label the irreducible representations of $S_N$.
\begin{defn}
A Young diagram for $S_N$ is a collection of $N$ boxes, arranged in successive rows, such that each row contains fewer boxes than the previous one.
\end{defn}
A Young diagram is associated with a partition with elements equal to the number of boxes in each row of the Young diagram; see Figure \ref{fig:youngdiagram} for a simple example. 

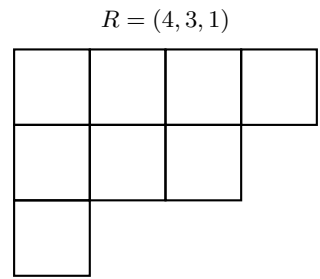
\begin{figure}[H]
    \centering
    \begin{tikzpicture}[scale=1]

      % Row 1: 4 boxes
      \foreach \col in {0,1,2,3} {
        \draw[thick] (\col, 0) rectangle ++(1,-1);
      }
      % Row 2: 3 boxes
      \foreach \col in {0,1,2} {
        \draw[thick] (\col, -1) rectangle ++(1,-1);
      }
      % Row 3: 1 box
      \draw[thick] (0, -2) rectangle ++(1,-1);

      % Label lambda above the diagram
      \node[above] at (2, 0.1) {$R = (4,3,1)$};

    \end{tikzpicture}
  \hfill
  \caption{The Young diagram associated with the partition $R = (4,3,1) \vdash 8$.}
           \label{fig:youngdiagram}
\end{figure}

\item[]{\bf Plancherel measure.} It is extremely convenient to consider a measure on the space of Young diagrams of $N$ boxes. The natural measure is called the Plancherel measure, and evaluated on an irreducible representation $R$ it is given by
\eq{
\mathrm{Pl}_N(R)=\frac{(\dim R)^2}{N!}\,.
}
In fact, it is easy to see that the Plancherel measure is a probability measure that can be defined on any finite group once we replace the factor $N!$ in the denominator with the order of the group. 

\item[]{\bf Poissonized Plancherel measure.} We can generalize the Plancherel measure to the grand-canonical setting \cite{Bogachev2006Central}. Evaluated on an irreducible representation $R$, labeled by a Young diagram of $\abs{R}$ boxes, it is given by
\eq{
\widetilde{\mathrm{Pl}}_{p}(R)=e^{-p}p^{\abs{R}}\left(\frac{\dim R}{\abs{R}!}\right)^2\,.
}
For our purposes, the Plancherel and Poissonized Plancherel measure are interchangeable in the limit $p\sim N\rightarrow\infty$ \footnote{This is true up to subleading corrections in the weak-coupling limit.}. This can be thought of as the equivalence of ensembles in the thermodynamic limit.

\item[]\textbf{Kerov's central limit theorem.} Here we simply state Kerov's central limit theorem, which is the main statement we reinterpret in terms of baby universe field theory. We denote by $\chi_R(k)$ the value of the character of an irreducible representation $R$ of $S_{|R|}$ on the single cycle $(1\dots k)$.
\begin{thm}[Kerov,\cite{ivanov1999algebra,ivanov2001algebra}]\label{thm:charclt}
Let $n\in\mathbb{N}$ be fixed. The functions $X_{k}(R)\coloneqq\frac{\abs{R}!}{(\abs{R}-k)!}\frac{\chi_R(k)}{\dim R},\ k\leq n$, with $R$ sampled according to the Plancherel measure, converge in distribution as $\abs{R}\rightarrow\infty$ towards independent Gaussian random variables. More precisely, for any set of independent, standard Gaussian random variables $\xi_1,\dots,\xi_n$ and for any bounded, continuous function $f:\mathbb{R}^n\rightarrow\mathbb{R}$,
\begin{align}
\begin{split}
&\underset{\abs{R}\rightarrow\infty}{\mathrm{lim}}\,\mathbb{E}_{\mathrm{Pl}_{\abs{R}}}
\left(f\left(\frac{X_{2}}{\abs{R}^{\frac{2}{2}}},\dots,
\frac{X_n}{\abs{R}^{\frac{n}{2}}}\right)\right.\\
&\hspace{80pt}\left.-f\left(\sqrt{2}\xi_2,\dots,\sqrt{n}\xi_n\right)\right)=0\,.
\end{split}
\end{align}
\end{thm}
\end{itemize}

\bibliography{ref,all}

\end{document}